\documentclass{IOS-Book-Article}

\usepackage{mathptmx}
\usepackage{soul}\setuldepth{article}
\usepackage{graphicx}
\usepackage{booktabs}
\begin{document}

\pagestyle{headings}
\pagenumbering{arabic}
\begin{frontmatter}              

\title{LLM-enabled Social Agents}


%

\markboth{}{July 2026}

\author[A]{\fnms{\"{O}nder} \snm{G\"{u}rcan}\orcid{0000-0001-6982-5658}
\thanks{Corresponding Author: \"{O}nder G\"{u}rcan, NORCE Research AS, Kristiansand, Norway. Email: ongu@norceresearch.no}}
and
\author[B]{\fnms{Moharram} \snm{Challenger}\orcid{0000-0002-5436-6070}}

\runningauthor{\"{O}. G\"{u}rcan et al.}

\address[A]{NORCE Research AS, Universitetsveien 19, Kristiansand, Norway. E-mail: ongu@norceresearch.no}

\address[B]{University of Antwerp and Flanders Make, Belgium. E-mail: moharram.challenger@uantwerpen.be}

%
%
%
%

\begin{abstract}
Large Language Models (LLMs) have transformed agent–agent and human–agent interaction by enabling software, physical, and simulation agents to communicate and deliberate through natural language. 
Yet fluent language use does not by itself yield socially intelligible behaviour. Most current systems remain weakly grounded in roles, norms, intentions, and contextual constraints, limiting their capacity for meaningful participation in social environments. 
This paper develops a conceptual baseline for LLM-enabled social agents by arguing that they should be grounded in role definitions operationalized through persona descriptions. 
On this basis, we outline research directions for representation, hybrid control, and evaluation. 
The paper concludes that persona-based role definitions are a necessary foundation for turning language competence into social behaviour.
\end{abstract}

\end{frontmatter}
\markboth{July 2026}{July 2026}

\section{Introduction}

LLM-enabled multi-agent systems (MAS) are increasingly embedded in everyday life as they expand their presence across a wide range of domains including personal assistants \cite{He2025,Chan2025,Li2024}, autonomous robotics \cite{Rafique2025,Liang2025}, autonomous driving, healthcare diagnostics \cite{Xu2025,Yang2025}, supply chains \cite{Zhao2025}, games and social simulations \cite{gurcan2025towards,Gurcan2024LLM,Borah2025}.
Despite this integration, many of these systems are still predominantly designed from a technocentric perspective. 
Hence, the design and development of these systems often prioritize technical capabilities and efficiencies (such as scalability, communication efficiency, and autonomous task decomposition) over social considerations (such as social enactment, fairness, explainability, trust, ethical guidelines and human-centered design),  which are critical for real-world deployment and societal acceptance \cite{Chowa2026,Shaikh2026,Gurcan2024LLM,Sissodia2025}.
Recent discourse and empirical studies highlight the necessity of balancing technical efficiency with social enactment, especially as LLM-MAS are increasingly deployed in domains impacting social welfare, civic engagement, and sensitive decision-making \cite{Zhang2025,Li2024}.

Large language models (LLMs) enable natural language understanding \cite{Shao2024,Barbon2024}, contextual interpretation \cite{Mohamed2025,Raiaan2024}, flexible response generation \cite{Shao2024}, and increasingly sophisticated forms of reasoning and planning \cite{Zhao2025b,Jiang2024}. 
In multi-agent systems, these capabilities make it possible to augment more traditional architectures with language-based deliberation, allowing agents to interpret instructions \cite{Xiang2025,Lee2025,Lim2024}, coordinate with other agents \cite{Yang2026,Gogineni2025,Chen2024}, interact with humans \cite{Lee2025,Guo2026}, and adapt their behaviour to changing contexts \cite{Shaikh2026}. 
Recent work on LLM-enabled and generative agents further suggests that memory \cite{Jia2025}, reflection \cite{Tan2025,Liang2025b,Ma2025}, and persona-related representations \cite{Tissaoui2026,park2023generative} can support more coherent and context-sensitive behaviour over time. 
At the same time, existing approaches still often privilege linguistic fluency and task completion \cite{Zhang2025b,Wu2025} over persistent intention management \cite{Zhang2025b,Savarimuthu2025}, explicit role grounding \cite{Takagi2025,Perera2025}, and social action \cite{Munz2026,Mou2025}.

This limitation becomes especially important when considering social agents. 
A social agent is not merely an entity that communicates in natural language or reacts to observable cues, but one that operates in relation to social roles, norms, expectations, relationships, and situated interaction contexts \cite{Foo2007,Jackson2021,Governatori2008,Fasli2004}. 
In contrast to generalized user modelling or interface-level personalization, social agency requires that social information be represented within the agent’s internal reasoning process itself, so that perception, interpretation, goal formation, and action selection are shaped by who the agent is, what role it performs, and what constraints and responsibilities follow from that role \cite{Dignum2015}. 
From this perspective, social cognition in artificial agents depends not only on interaction capability, but also on structured representations that make behaviour understandable, consistent, and socially meaningful across time and situations \cite{Hortensius2018,Lukasik2025}.

The novelty of this paper is conceptual rather than empirical. We argue that persona descriptions should be treated as the representational baseline for LLM-enabled social agents (Section \ref{sec:The-Conceptual-Baseline-for-LLM-enabled-Social Agents}), and we use this claim to motivate a research agenda on representation, control, and evaluation (Section \ref{sec:Research-Directions}).

\section{The Conceptual Baseline for LLM-enabled Social Agents}
\label{sec:The-Conceptual-Baseline-for-LLM-enabled-Social Agents}

A conceptual baseline is needed to clarify what should count as an LLM-enabled social agent before discussing architectures, methods, or application domains. 
Current LLM-enabled agent systems predominantly integrate natural language interaction with task execution, yet they largely treat sociality as an external add-on instead of embedding it as a foundational aspect of the agent's cognitive model  \cite{Mou2025,Gurcan2024LLM}.  
This limits the system's capacity to interpret intentions, maintain context-sensitive behaviour, and reason about interaction in socially meaningful ways. 
A stronger baseline is therefore required, one in which sociality is not added after deliberation, but built into the agent definition itself.

To justify any baseline for LLM-enabled social agents, it is necessary to consider the criteria such a baseline should satisfy. 
A useful baseline should be interpretable, so that role-related assumptions remain explicit and inspectable; action-guiding, so that it can meaningfully shape agent behaviour rather than merely label it; compatible with LLM-based deliberation, so that it can be naturally incorporated into language-mediated reasoning; persistent across time, so that the agent can maintain behavioural coherence across interactions; supportive of norm integration, so that social expectations, obligations, and constraints can be attached to the role representation; and sufficiently simple to evaluate, so that alternative implementations can be compared empirically.

From this perspective, an LLM-enabled social agent should not be understood merely as an agent that can communicate fluently in natural language. Linguistic fluency is not sufficient for social meaning. 
A social agent should be able to act in relation to roles, expectations, constraints, and situated interaction contexts \cite{Sanders2014}. 
Its decisions should not be generated only from immediate prompts or isolated goals, but from an internally meaningful representation of who the agent is, what role it currently performs, what boundaries and responsibilities follow from that role, and how its behaviour should remain interpretable over time. 
Hence, we define an LLM-enabled social agent as a role-enacting agent whose role is operationalized through a persona description that grounds deliberation in traits, values, goals, constraints, relationships, and commitments, and whose context-sensitive behaviour is generated through LLM-centered deliberation, complemented where needed by additional mechanisms for memory, control, and normative regulation.

\begin{figure}[th]
\centering
\includegraphics[width=\textwidth]{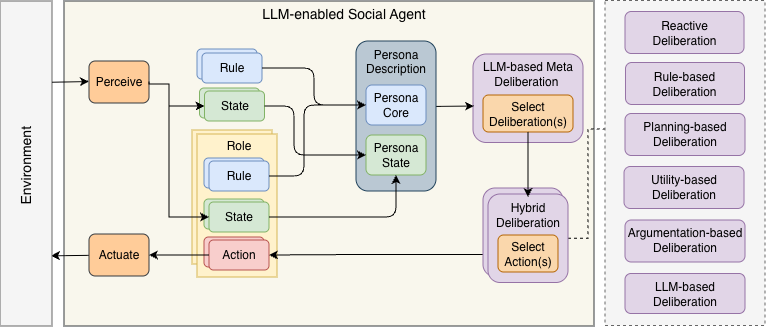}
\caption{An LLM-enabled Social Agent's Internal Architecture (improved from \cite{Asici2026})}
\label{fig:social-agent}
\end{figure}

The key conceptual step is to define LLM-enabled social agents through roles rather than generic capability bundles. 
In social environments, an agent is a situated participant that occupies one or more roles, and these roles shape what it treats as relevant, what it prioritizes, what actions are legitimate, and how its behaviour is interpreted by others. 
However, role labels alone are too weak to support social behaviour. A label such as “caregiver” or “assistant” does not by itself specify how the agent should interpret situations, resolve trade-offs, or sustain behavioural consistency across interactions. 
For this reason, roles should be operationalized through \textit{persona descriptions}, which make them computationally meaningful for LLMs by specifying traits, values, goals, needs, skills, constraints, resources, boundaries, and interaction tendencies.
Note that, persona descriptions are not understood merely as free-form natural-language prompts. 
Rather, they should be seen as hybrid symbolic-textual representations: partially readable as natural-language descriptions, but organized into explicit fields such as role, stable traits, values, goals, responsibilities, boundaries, dynamic state, norms, relationships, and commitments.
In more advanced implementations, such persona representations may also be partially updated or dynamically learned as internal models through interaction, memory, and adaptation over time.

As illustrated in Figure \ref{fig:social-agent}, an LLM-enabled social agent is grounded in a persona-based role representation, while LLM-based meta deliberation remains central but is complemented by memory, normative structures, and control mechanisms, , following the broader direction of hybrid role-based architectures for LLM-enhanced multi-agent systems \cite{Asici2026}.
The persona description can be understood as comprising two complementary parts: a Persona Core and a Persona State. 
The Persona Core contains the relatively stable elements that preserve the agent’s social identity across situations, such as role grounding, enduring traits, values, boundaries, responsibilities, and longer-lived commitments. 
The Persona State contains the more dynamic elements that change through interaction and context, such as current interpretations, situational priorities, relationship status, and memory-informed updates. This distinction helps clarify how persona descriptions can remain stable enough to preserve behavioural coherence while also adapting to interaction history, situational change, and environmental feedback.
%
%

This baseline also implies that persona descriptions should be dynamically assembled rather than treated as fixed prompts: stable enough to preserve identity and behavioural coherence, yet flexible enough to adapt to memory, interaction history, situational change, and environmental feedback.
%
Within this baseline, LLMs provide the main interpretive layer: they map persona-rich role representations onto context-sensitive understanding and action. 
However, this does not make the architecture LLM-only, nor does it require the LLM to internally remember all evolving social context. Instead, memory systems, relationship models, commitment stores, normative structures, and environmental state can maintain dynamic contextual information outside the LLM. At each deliberation step, the agent can inject a selected, up-to-date representation of this information into the Persona State. The baseline therefore treats long-term social coherence as an architectural property of the agent, not as a capability delegated entirely to the LLM.
This baseline also clarifies the distinction between generic LLM agents and LLM-enabled social agents. A generic LLM agent may answer, plan, retrieve, or converse effectively without a stable social identity. A social agent, by contrast, must enact a role through a persona description and maintain coherence between persona, context, and behaviour.

Note that, we do not claim that persona descriptions are sufficient for social agency. 
Rather, we claim that they provide an appropriate representational baseline for LLM-enabled social agents, because they make roles behaviorally meaningful in a form that LLMs can operationalize. Norms, obligations, relationships, institutional roles, and memory remain indispensable complementary structures.
Persona descriptions should therefore not be understood as isolated character sketches. To support social deliberation, they should be enriched with complementary social structures, including norms, institutional roles, relationship models, and commitments. 
Norms specify what is permitted, expected, or forbidden; institutional roles specify formal responsibilities and authority relations; relationship models capture interactional asymmetries and mutual expectations between agents; and commitments represent explicitly undertaken obligations that persist across interactions. 
In this way, persona descriptions provide the internal behavioural grounding of the role, while these additional structures provide its normative, relational, and institutional embedding.

Several alternative baselines could be used to ground social agency, including role labels, normative rule sets, institutional role models, and relational or commitment-based representations (Table~\ref{tab:baseline-comparison}). 
These alternatives are valuable, but they serve different representational functions. 
Role labels alone remain too abstract to guide situated behavior, since they specify a category without sufficiently constraining its interpretation in context. 
Normative rule sets capture obligations and prohibitions, but they primarily encode external constraints rather than an internally usable behavioral orientation. 
Institutional role models formalize positions, permissions, and responsibilities within structured organizations, yet they are less effective as a general baseline for open-ended social interaction. 
Relational and commitment-based models represent social dependencies and expectations between agents, but they typically require a richer surrounding social specification and do not by themselves furnish an immediately actionable deliberative perspective. 
In comparison, persona descriptions provide a more suitable baseline for LLM-based deliberation because they render roles behaviorally interpretable in a form that LLMs can directly operationalize. 
A persona can compactly express motivations, dispositions, communicative style, contextual expectations, and practical priorities, thereby supplying the LLM with a socially meaningful deliberative frame. 
For this reason, we argue that personas constitute the most appropriate representational foundation, to be complemented by norms, institutional structure, relationships, commitments, and memory.

\begin{table}[t]
\centering
\caption{Comparison of alternative conceptual baselines for LLM-enabled social agents.}
\label{tab:baseline-comparison}
\begin{tabular}{p{2.7cm}p{4.2cm}p{4.6cm}}
\toprule
\textbf{Baseline} & \textbf{Strength} & \textbf{Limitation} \\
\midrule
Role label &
Simple, compact, and easy to interpret &
Too abstract to guide situated, socially coherent behaviour \\

Norm rules &
Makes obligations, permissions, and prohibitions explicit &
Weak at capturing internal orientation, style, and open-ended behavioural variation \\

Institutional role model &
Provides strong formal structure and clear organizational grounding &
Less suited to flexible, naturalistic, open-ended interaction \\

Relationship / commitment model &
Captures social dependencies, expectations, and inter-agent obligations well &
Requires broader specification and may be cumbersome as a minimal baseline \\

Persona description &
Behaviorally interpretable for LLMs and expressive enough to ground role enactment &
Not sufficient on its own without complementary norms, memory, and relational structure \\
\bottomrule
\end{tabular}
\end{table}

This reveals what the field is currently missing: a shared baseline that makes social roles computationally actionable for LLM-based agents. On that basis, the next section proposes a concrete agenda shift toward representation, control, and evaluation of persona-grounded social agents.


%

\section{Research Directions}
\label{sec:Research-Directions}

Given this conceptual baseline, three research challenges follow: how persona-grounded social agents should be represented (Section \ref{sec:Representation}), how role-consistent behaviour should be preserved through LLM-centred but hybrid deliberation (Section \ref{sec:Control}), and how such agents should be evaluated against weaker alternatives (Section \ref{sec:Evaluation}).


\subsection{Representation: From Role Labels to Persona-Grounded Social Models}
\label{sec:Representation}

A first research direction concerns how persona-grounded role representations should be specified, organized, and updated. As argued in the previous section, role labels alone are too abstract to guide situated behavior. Labels such as teacher, caregiver, or manager indicate a broad social function, but they do not specify how an agent should interpret priorities, negotiate expectations, manage tensions, or respond to contextually varied situations. Persona descriptions provide a stronger baseline because they make roles behaviorally interpretable for LLM-based deliberation by encoding relevant dispositions, tendencies, preferences, commitments, boundaries, and styles of interaction.

Future work should therefore investigate how persona descriptions can be designed as structured but flexible representations rather than as informal prompt fragments or static character sketches. In particular, persona-grounded social agents require representational models that preserve coherence across situations while remaining responsive to interaction history, institutional context, and environmental change. This raises questions about which dimensions of persona are most consequential for social behavior, how persona elements should be organized, which aspects should remain stable, and which should be dynamically updated over time.
%
A social agent must also be grounded in norms \cite{Viana2015,DosSantosNeto2012}, commitments \cite{FernandezCastro2023,Carabelea2006}, institutional expectations \cite{Fasli2006}, and relational context \cite{Dimas2013}. 
The challenge is therefore not only to describe who the agent is in persona terms, but also to represent what the agent owes to others, what social constraints shape its actions, and how its role is situated within a broader social and organizational environment. Future research should examine how textual persona descriptions can be complemented by more explicit representations of social relationships, normative structures, role obligations, and contextual priorities without losing the expressive advantages of LLM-based interpretation.

The field also needs \textit{synthetic persona datasets} as a shared experimental infrastructure for LLM-enabled social agents. 
Machine learning advanced rapidly once common datasets made systematic comparison possible \cite{Nguyen2019}; social-agent research remains largely in a pre-benchmark stage, dominated by bespoke prompts and one-off demonstrations. 
Without shared persona datasets, it is difficult to determine whether apparent social competence reflects genuine role grounding or merely prompt-specific performance. Carefully designed synthetic persona datasets could encode controlled variation in roles, values, commitments, relationships, cultural settings, institutional constraints, and memory traces, thereby enabling rigorous evaluation of consistency, norm handling, social adaptation, and longitudinal coherence. More importantly, they would allow the field to accumulate knowledge: not just whether a system works in one crafted scenario, but which representational ingredients and architectural choices reliably produce social behavior across settings. Synthetic persona datasets should therefore be treated as a foundational research priority rather than a secondary tooling issue.

Such datasets would also help scale persona-grounded agent research beyond a handful of manually authored examples. They could enable experiments in which one dimension is varied while others are held constant, making it easier to study the contribution of specific persona elements to behavioral coherence, role fidelity, or social appropriateness. At the same time, the creation of synthetic persona datasets is itself a research challenge. These datasets may encode cultural narrowness, stereotypical assumptions, or artificial regularities that fail to reflect the complexity of real social life. For that reason, future work should address not only how to use such datasets, but also how to construct them transparently, diversify them responsibly, and document their limitations.

\subsection{Control: LLM-Centred but Hybrid Deliberation}
\label{sec:Control}

A second research direction concerns deliberation and execution. Within the proposed baseline, LLMs remain central because they provide the interpretive flexibility needed for situated social reasoning, contextual response generation, and adaptive interaction. However, they do not by themselves guarantee persistence of commitments, long-term coherence, reliable planning, or stable enactment of persona-grounded roles. If persona descriptions are to function as more than descriptive overlays, future research must investigate how social agents can preserve persona-grounded behavior throughout deliberation, memory use, and action execution. As suggested by Figure 1, this should be understood not as purely LLM-only deliberation, but as LLM-centred hybrid deliberation, in which the LLM remains the main interpretive mechanism while other deliberative and control processes are composed around it.

This suggests the importance of hybrid architectures in which LLM-based deliberation is complemented by additional mechanisms for memory, planning, constraint handling, and action control \cite{Asici2026}. The key issue is not whether symbolic or sub-symbolic approaches should dominate, but how complementary mechanisms can support the continued enactment of persona-grounded roles without displacing the LLM’s central role in interpretation and socially situated reasoning. 
More specifically, hybrid deliberation should be understood as a compositional deliberative scheme: it can incorporate and coordinate multiple forms of deliberation, such as reactive deliberation \cite{Navarro2009}, rule-based or normative deliberation \cite{Chien1991}, planning-based deliberation \cite{Alford2016,Debenham2008}, and utility-based deliberation, argumentation-based deliberation \cite{Kok2011,Tang2005}, or LLM-based deliberation \cite{Yu2025}. 
In this view, hybrid deliberation is not merely the coexistence of modules, but an orchestration layer that selects, combines, constrains, or sequences different deliberative modes depending on the current context, persona state, and role obligations. Such architectures may therefore include explicit memory systems for preserving commitments and relational history, planning modules for maintaining longer-term consistency, reactive mechanisms for timely situated response, and constraint mechanisms for enforcing normative or institutional boundaries.

One particularly important challenge is the coordination of short-term and long-term deliberation. 
Social agents must often respond to immediate conversational or situational demands while also remaining faithful to enduring commitments \cite{Flores2004}, role expectations \cite{Galland2022,Ghosh2021}, and accumulated interaction history. 
A persona-grounded agent should therefore not only produce locally plausible responses, but also sustain a recognizable pattern of conduct across time. 
Future work should examine how architectures can maintain continuity between momentary judgments and longer-term persona-grounded role enactment, especially in settings where social expectations evolve or conflict.

Memory is central here. Without mechanisms for preserving relevant experiences, commitments, relational trajectories, and prior decisions, persona-grounded behavior risks collapsing into episodic improvisation. 
Yet memory should not be treated merely as a storage layer. 
In social agents, memory plays a constitutive role in sustaining who the agent appears to be across repeated interactions \cite{Campos2018,Richards2014,Aylett2013}. 
This raises questions about what should be remembered, how memories should be selected or summarized, how they interact with persona descriptions, and how agents should balance persistent identity with contextual adaptation.

\subsection{Evaluation: Measuring Social Intelligibility}
\label{sec:Evaluation}

A third research direction concerns evaluation. If persona-grounded role definitions are to serve as a meaningful conceptual baseline, they must be evaluated not only as an appealing design intuition but as a stronger foundation than weaker alternatives such as generic LLM agents, role labels alone, or externally imposed rule sets. Future work should therefore develop evaluation frameworks that go beyond task completion and linguistic fluency to assess whether agents behave in social ways.

Relevant evaluation criteria include behavioral coherence across interactions, fidelity to persona and role, persistence of commitments, sensitivity to relational and normative context, and interpretability of behavior to human users or observers. 
These dimensions are especially important in domains where agents are expected to participate in socially meaningful interaction (such as social simulations \cite{Gurcan2024LLM}) rather than merely provide information or execute isolated tasks. Evaluation should also examine whether persona-grounded agents remain coherent under contextual variation, conflicting expectations, incomplete information, or extended interaction sequences.

Shared synthetic persona datasets would be especially valuable in this context because they could support benchmark-style evaluation \cite{Shaheen2024}. 
By providing standardized but diverse persona configurations, they would enable systematic comparison across agent architectures, prompting strategies, memory mechanisms, and control frameworks. 
They would also make it possible to conduct ablation-style studies, for example by comparing agents grounded in rich persona descriptions with agents specified only through role labels, or by examining how the addition of commitments, norms, or relationship models affects social behavior under controlled conditions. 
In this way, synthetic persona datasets would support reproducibility, comparability, and cumulative empirical progress in a field that currently relies too heavily on handcrafted examples and anecdotal demonstrations.

Evaluation should also include the study of failure modes. Persona-grounded agents may still drift, overfit to superficial cues, become inconsistent over time, or reproduce biased assumptions embedded in their persona specifications or training materials. For this reason, future work should investigate not only success criteria but also mechanisms for diagnosing representational weakness, persona instability, normative breakdown, and socially unintelligible responses. A mature research programme will require methods for identifying where persona-grounded models fail, why they fail, and how such failures differ from those observed in simpler baselines.

Taken together, these research directions suggest that persona-grounded social agency should be pursued not merely as a prompting technique or implementation detail, but as a broader design and evaluation paradigm. The field now needs richer representations, hybrid control mechanisms, synthetic persona datasets, and comparative benchmarks that together can support the systematic development of socially coherent LLM-enabled agents.

\section{Conclusion}
We argued that LLM-enabled social agents should be understood as role-playing agents whose roles are operationalized through persona descriptions and enacted through LLM-centered deliberation, complemented by normative, relational, and operational mechanisms. This framing provides a clearer conceptual foundation for designing and evaluating social agent systems. Future work should test this baseline empirically against alternative role-grounding approaches.

\bibliographystyle{vancouver}   
\bibliography{bibtex}

\end{document}